\documentclass[showpacs,twocolumn,preprintnumbers,amsmath,amssymb,prl]{revtex4}

\usepackage{graphicx}% Include figure files
\usepackage{dcolumn}% Align table columns on decimal point
\usepackage{bm}% bold math

\begin{document}

%\preprint{APS/123-QED}

\title{Demonstration of a Controllable Three-Dimensional Brownian Motor \\in Symmetric Potentials}

\author{P. Sj\"olund$^{1}$}
\author{S. J. H. Petra$^{1}$}
\author{C. M. Dion$^{1}$}
\author{S. Jonsell$^{1}$}
\author{M. Nyl\'en$^{1}$}
\author{L. Sanchez-Palencia$^{2}$}
\author{A. Kastberg$^{1}$}%
 \email{anders.kastberg@physics.umu.se}
\affiliation{%
$^{1}$Department of Physics, Ume\aa \ University, SE-90187 Ume\aa , Sweden \\
$^{2}$Laboratoire Charles Fabry de l'Institut d'Optique,
CNRS and Universit\'e Paris-Sud XI,
B\^at 503, Centre scientifique, F-91403 Orsay cedex, France
}

\date{\today}

\begin{abstract}
We demonstrate a Brownian motor, based on cold atoms in optical lattices, where isotropic random fluctuations are rectified in order to induce controlled atomic motion in arbitrary directions. In contrast to earlier demonstrations of ratchet effects, our Brownian motor operates in potentials that are spatially and temporally symmetric, but where spatiotemporal symmetry is broken by a phase shift between the potentials and asymmetric transfer rates between them. The Brownian motor is demonstrated in three dimensions and the noise-induced drift is controllable in our system.
\end{abstract}

\maketitle

%Introduction
Understanding how to extract useful work out of the energy of random fluctuations in mesoscopic or quantum mechanical systems is a formidable challenge from a practical as well as from a conceptual viewpoint. On one hand, this may have an impact on nanotechnology, and 
also on physiology since many protein motors, responsible for intra-cell transport, work this way~\cite{Oster:2003}. On the other hand, the realization of a Brownian motor (BM) faces the inherent difficulty that one has to circumvent two fundamental physical principles. Firstly, from the Curie principle~\cite{Curie:1894}, it follows that no directed motion can emerge in the absence of an asymmetry. Secondly, converting the energy of a thermal bath into deterministic work would be ruled out by the second law of thermodynamics. Therefore, designing a BM requires two necessary conditions, namely that the system must be (i) spatio-temporally asymmetric and (ii) brought out of thermodynamic equilibrium. Although difficult to prove, it is believed that under these constraints, the system will indeed realize a BM~\cite{Reimann:2002_PR}. The archetype proposals of BM's rely on the application of a force asymmetric in space or in time, albeit one that averages to zero.

The earliest example of a BM is the Smoluchowski-Feynman ratchet~\cite{Smoluchowski:1912,Feynman:1963}. A wide variety of BM's have since been theoretically investigated 
and also demonstrated for various systems (see recent reviews~\cite{Hanggi:2005_AnnPh,Astumian:2002,Reimann:2002_APA,Reimann:2002_PR} and references therein).
However, most of the realizations of BM's lack the possibility of inducing motion in three dimensions and are difficult to control. It has been proposed~\cite{Robilliard:1999} that the physics of BM's may be revisited in periodic arrays of ultra-cold atoms (optical lattices, OL's). These prove to be highly controllable and versatile systems~\cite{Jessen:1996,Grynberg:2001}.
In~\cite{Robilliard:1999}, Robilliard {\it et al.}\/ demonstrated a flashing ratchet in a periodic but spatially asymmetric OL\@. Directed motion in 1D has also been induced in a periodic and spatially 
symmetric OL, and in atom traps, where the source of asymmetry was an unbiased temporally asymmetric force~\cite{Schiavoni:2003,Jones:2004,Gommers:2005_PRL1,Gommers:2005_PRL2,Dykman:1997}. 

More subtle ratchet effects can arise in the absence of spatial or temporal asymmetry of the trapping potentials provided that unbiased non-equilibrium perturbations induce a dynamical spatial asymmetry and thus break the detailed balance~\cite{Luczka:1995,Hanggi:1996,Reimann:2002_PR}. It has been suggested that this can be achieved when asymmetric switching 
rates between two symmetric potentials are combined with a spatial shift of these potentials~\cite{Sanchez-Palencia:2004}. In this case, the spatio-temporal symmetry is broken, even though the potentials used are symmetric.

In this Letter, we report the first demonstration of a BM based on such a mechanism. It consists of ultra-cold atomic gases switching between \emph{two} state-dependent OL's, both spatially and temporally symmetric, that are coupled via optical pumping. The rectification process emanates from the fact that  (i) the two OL's are spatially shifted and (ii) the couplings between the two potentials, via the vacuum field reservoir, are strongly asymmetric. The spatial shift between the OL's is adjusted at will and the transition rates can be controlled via the frequency and intensity of the OL lasers. A directed motion at constant velocity is obtained except for specific parameters where the detailed balance is not broken. Moreover, the directed motion can be induced in any direction in three dimensions. This new type of a BM opens up possibilities for fundamental studies of noise-induced directed motion. The underlying principle is very general and is potentially transferable to molecular motors and to applications in nano-electronics and chemistry.

\begin{figure}
\begin{center}
\includegraphics[scale=0.4]{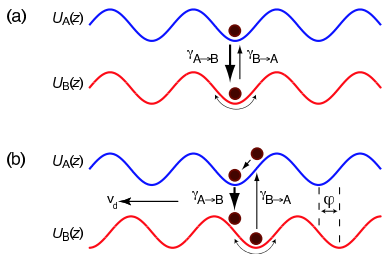}
\caption{Rectification mechanism. Atoms move in two symmetric potentials $U_{\mathrm{A}}(z)$ and $U_{\mathrm{B}}(z)$ that are coupled via the asymmetric optical pumping rates $\gamma_{ \mathrm{A} \rightarrow  \mathrm{B}}$ and $\gamma_{ \mathrm{B} \rightarrow  \mathrm{A}}$ ($\gamma_{ \mathrm{A} \rightarrow  \mathrm{B}} \gg \gamma_{ \mathrm{B} \rightarrow  \mathrm{A}}$). a) The potentials are in phase. The transfer from the long lived state B, to the transient state A, and back, will not lead to biased motion. b) A phase shift $\varphi$ is introduced. Spatial diffusion will be strongly facilitated in one direction, leading to a drift velocity $v_\mathrm{d}$. \label{rectificationmech}}
\end{center}
\end{figure}

%Rectification mechanism
The basic idea for the rectification mechanism is depicted in Fig.~\ref{rectificationmech}. Depending on their internal state, the atoms are subjected to one of two three-dimensional periodic potentials ($U_\mathrm{A}$ and $U_\mathrm{B}$) with identical periods. The atoms have low kinetic energy compared to the modulation depth of the potentials and they undergo Hamiltonian motion around potential minima. This is interrupted by a dissipative process, which eventually causes a Brownian diffusion in the potential. The dissipation also means that an atom can be pumped from one internal state to the other, resulting in random jumps between $U_\mathrm{A}$ and $U_\mathrm{B}$. The asymmetry that eventually gives rise to rectification is caused by a pronounced difference in the transfer rates between the potentials
($\gamma_\mathrm{A \rightarrow B} \neq \gamma_\mathrm{B \rightarrow A}$). If the two potentials are in phase (Fig.~\ref{rectificationmech}a), an atom will spend most of its time in the long lived state (B), where its dynamics can be well described by a quantum-mechanical harmonic oscillator with dissipation. Every now and then, it is pumped to the transient state (A), from where it returns very quickly to state B. This excursion will cause a slight heating and will increase the probability for escape to a neighboring lattice site. This diffusion is symmetric. If the the potentials are shifted (Fig.~\ref{rectificationmech}b), the situation changes drastically. During the time spent in lattice A, the atom experiences a potential with an incline that depends on the phase shift. Thus, the diffusion will be strongly enhanced in one specific direction, and correspondingly reduced in the opposite direction. Even though the potentials are symmetric and stationary, the atoms are propelled in a controllable direction.

%%%%%%%%%%%%%%%%%%%%%%%%%%%%%%%%%%%%%%%%%%%%
%%%%%%%%%%%%%%%%%%%%%%%%%%%%%%%%%%%%%%%%%%%%

%Optical lattices
We realize such a BM, in a controllable fashion, using cold atoms in a double OL\@. Optical lattices are spatially periodic potentials formed in the interference pattern from a number of laser beams due to a second-order interaction between an atomic dipole and the light field~\cite{Jessen:1996,Grynberg:2001}. Tuning the light frequency of an OL to the proximity of an atomic resonance 
provides a dissipative channel resulting from spontaneous emission. The potential is then 
accompanied by an efficient cooling mechanism~\cite{Dalibard:1989,Chu:1998,Cohen-Tannoudji:1998,Phillips:1998}, providing friction in  our system, and by momentum diffusion leading to a Brownian motion of the atoms~\cite{Dalibard:1989,Sanchez-Palencia:2002}. The presence of dissipation, albeit small, will result in a slow normal diffusion of the atoms~\cite{Hodapp:1995,Grynberg:2001,Sanchez-Palencia:2002}. The specificity of our set-up is that it consists of \emph{two} OL's produced by two different laser fields of slightly different frequencies, tuned close to distinct atomic resonances. This is realized using cesium, which has a hyperfine structure in the ground state that is great enough to enable spectrally selective OL's, but also small enough to ensure that the two distinct lattices have essentially the same spatial periodicity (given by the wavelengths of the light and the interference pattern) within the sample volume.  In these double OL's, optical pumping also induces asymmetric couplings between the potentials and it is possible to accurately control the relative spatial phase between the potentials as demonstrated in~\cite{Ellmann:2003_PRL,Ellmann:2003_EPJD}.

This physical set-up is not an exact replica of the simple model depicted in Fig.~\ref{rectificationmech} but yields a system that shares the same basic features. Instead of the two potentials, $U_{\mathrm{A}}(z)$ and $U_{\mathrm{B}}(z)$, the atoms now shift between two manifolds of potentials, corresponding to two different hyperfine levels in the ground state of Cs. Optical pumping will preferentially transfer atoms to the magnetic substates with the deepest optical potentials. Apart from transition between the manifolds, there will also be transitions within the respective manifolds. In fact, transitions of this kind are responsible for the dissipation~\cite{Dalibard:1989,Grynberg:2001,Dion:2005}. This dissipation is state and position dependent, but can, by spatial averaging, be likened to a friction force and a momentum diffusion tensor~\cite{Dalibard:1989,Castin:1991}.

%experimental setup
The experimental apparatus has been described in detail elsewhere~\cite{Ellmann:2003_PRL,Ellmann:2003_EPJD,Petra:2005}. In brief, we start with a cloud of laser-cooled Cs atoms with a temperature of a few microkelvin. Three-dimensional, tetragonal OL's are formed by intersecting sets of four laser beams. Two of these beams propagate in the $xz$-plane, with an angle of 45$^\circ$ with the $z$-axis, and are polarized along $y$. The other two beams propagate in a similar fashion in the $yz$-plane and are polarized along $x$. In order to create a double lattice (two overlapped OL's with the same spatial period and the same topography), two such sets are overlapped. When doing this, great care has to be taken to maintain phasestability and to avoid spurious drifts due to unbalanced radiation pressure~\cite{Petra:2005}. In the double OL, we typically trap some $10^8$ atoms, with a filling fraction of about 0.05 atoms per site. The atomic states trapped are the $F_\mathrm{g}=3$ and $F_\mathrm{g}=4$ hyperfine structure levels of the ground state in Cs (6s~$^2$S$_{1/2}$). The light fields are close to the D2 resonance at 852 nm (reaching 6p~$^2$P$_{3/2}$). Among the hyperfine structure manifolds in the ground and excited states, lattice B is close to the ($F_\mathrm{g}=4 \rightarrow F_\mathrm{e}=5$) resonance. This is a closed transition and the optical pumping out of lattice B will be very slow. In contrast, lattice A operates on the open transition ($F_\mathrm{g}=3 \rightarrow F_\mathrm{e}=4$), where the probability for optical pumping to lattice B is large. This provides the required asymmetric transition rates between lattices A and B. By varying the irradiances and the detunings from the two atomic resonances, the optical pumping rates can be controlled individually. In OL's, the origin of the lattice depends on the relative phases of the four laser beams~\cite{Grynberg:2001}. The spatial shifts between the two OL's along the three coordinate axes can thus be controlled individually by adjusting the optical path lengths of the four branches that build up the lattices \cite{Ellmann:2003_PRL,Ellmann:2003_EPJD}. 

%%%%%%%%%%%%%%%%%%%%%%%%%%%%%%%%%%%%%%%%%%%%
\begin{figure}
\begin{center}
\includegraphics[scale=0.45]{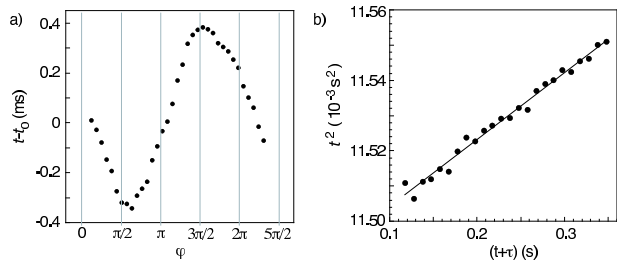}
\caption{a) Induced directed drift in the vertical ($z$) direction, measured as arrival time $t$ at a probe beam as a function of the relative spatial phase $\varphi$ for an interaction time $\tau$ of 350 ms. $t_0$ is the arrival time for zero drift velocity. b) Confirmation of a constant drift velocity obtained by plotting $t^2$ as a function of $t+\tau$ for different $\tau$. The line is a linear fit. \label{exp_tof}}
\end{center}
\end{figure}
%%%%%%%%%%%%%%%%%%%%%%%%%%%%%%%%%%%%%%%%%%%%

%directed motion
Experimental results are firstly shown in Fig.~\ref{exp_tof}a. This shows induced drift, when a spatial phase difference is imposed between the lattices. The data are acquired with a ballistic time-of-flight technique~\cite{Phillips:1998} (standard for experiments with ultra-cold atoms), where the atoms are dropped, when the lattices are abruptly switched off, after a selectable interaction time ($\tau$) and detected 5 cm lower. The arrival time will depend on the combination of the vertical ($z$) position and velocity of the atoms at the time of release. Clearly, the larger the velocity in the upward (downward) direction, the latter (sooner) the atoms will arrive at the detection altitude. The graph shows arrival time ($t$) as a function of relative phase ($\varphi$). An independent measurement of the phase shift is provided by a simultaneous measurement of the temperature as described in~\cite{Ellmann:2003_PRL,Ellmann:2003_EPJD}. There is no induced drift when the relative spatial phase is 0, $\pi$ or $2\pi$: these are positions where no biased motion is expected (see Fig.~\ref{rectificationmech}a). For all other phase shifts, there is clear evidence of induced drift as expected from detailed balance breaking (see Fig.~\ref{rectificationmech}b). The drift has extrema around $\pi/2$ and $3\pi/2$, with opposite signs. By controlling the lattice parameters, such as the optical pumping rates between the potentials, the potential depths, and also the dissipation in the system, we are able to influence the magnitude of the directed transport. We can induce drifts of the order of velocities corresponding to one atomic recoil, which is about 3~mm/s for a Cs atom scattering a lattice photon.

%constant velocity
A central feature of the rectification mechanism is that it should induce a constant drift velocity of the atoms~\cite{Sanchez-Palencia:2004}. To confirm constant velocity in the $z$ direction, we used a relative spatial phase of $3\pi/2$ and varied the interaction time $\tau$ (i.e., the time duration during which the atoms are left in the double OL) in the lattices from 10 ms to 350 ms. Using a simple dynamics, it is straightforward to show that for an atom having a constant velocity, $v_\mathrm{d}$, in a controllable time interval, $\tau$, before ballistic release, the arrival time, $t$, at the probe should have the functional form
\begin{equation}
gt^2+2v_\mathrm{d}(t+\tau)-2z_0=0  ,
\label{eq:constdrift}
\end{equation}
where $g$ is the gravitational acceleration and $z_0$ is the distance from the OL to the probe. In Fig.~\ref{exp_tof}b, we show that the data yield a linear relationship between $t^2$ and $(t+\tau)$. This confirms the assumption of constant velocity. In summary, our results demonstrate the BM described above that corresponds to inducing a directed drift at constant velocity in a set of two spatially shifted symmetric potentials with asymmetric transition rates.

%%%%%%%%%%%%%%%%%%%%%%%%%%%%%%%%%%%%%%%%%%%%
\begin{figure}
\begin{center}
\includegraphics[scale=1.1]{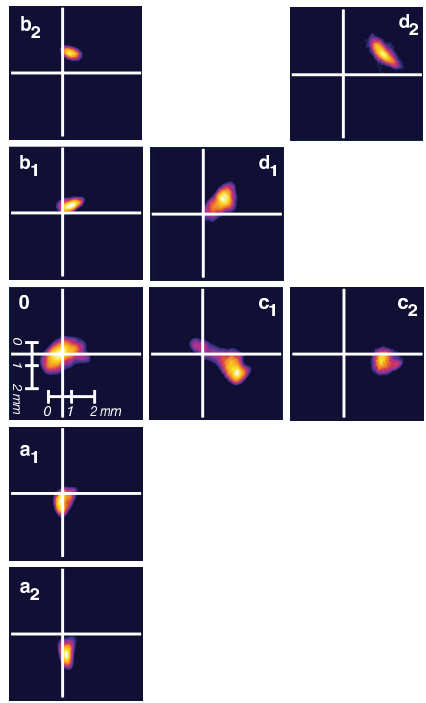}
\caption{Images of the atomic cloud. 
The image labeled `0' shows the initial cloud. 
The series labeled `a$_1$-a$_2$' are images taken at two different interaction times (0.4 and 0.9~s respectively) for a phase shift along $z$. For images `b$_1$-b$_2$', the phase shift is still along $z$, but differ by $\pi$. The series `c$_1$-c$_2$' is obtained when the phase shift is along $x$. Finally, a diagonal drift is displayed in `d$_1$-d$_2$', where the phase is shifted both along $z$ and along $x$. \label{images}}
\end{center}
\end{figure}
%%%%%%%%%%%%%%%%%%%%%%%%%%%%%%%%%%%%%%%%%%%%

\begin{figure}
\begin{center}
\includegraphics[scale=0.5]{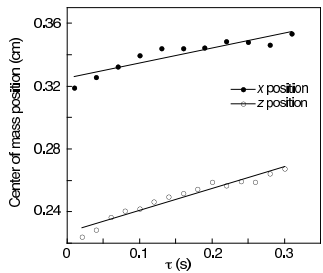}
\caption{Center-of-mass positions as a function of the interaction time $\tau$ in the OL for induced drift along $x$ and $z$. The data are obtained from images like those displayed in Fig.~\ref{images}. The lines are linear fits to the data\label{com_position}.}
\end{center}
\end{figure}

%In 3D
Our lattice structures are periodic in three dimensions, and we can adjust the relative spatial phases along $x$, $y$ or $z$ at will. Thus, the Brownian motor works also horizontally ($x$ and $y$ directions), and indeed in an arbitrary direction. Due to the absence of gravity in the horizontal directions, we have to use a different detection technique to confirm this. We measure directly the position of the center of the atomic cloud, as a function of interaction time $\tau$. This is done for horizontal direction $x$, as well as along $z$, by imaging the shadow of the atoms, transiently illuminated with a weak resonant probe beam, on a CCD-detector. The results are shown in Fig.~\ref{images} for the relative spatial phases that induce maximum drift in the respective directions. The data in Fig.~\ref{images} prove that the BM works also horizontally. To determine the evolution of center-of-mass position, we performed Gaussian fits of the atomic distribution. This method of detection is more direct than the time-of-flight method, but it has substantially lower resolution. Along $z$, the direct spatial data do, however, reproduce the results in Fig.~\ref{exp_tof}a. In Fig.~\ref{com_position}, we show the center-of-mass position measured as a function of interaction time $\tau$, yet again confirming that the directed drift has a constant velocity in both the $x$ and $z$ directions.

In our current set-up, we control the transition rates between the two potentials by varying the irradiance and the detuning of the light that builds up the potentials. The control is complicated by the fact that these modulations will also change the dissipation and the height of the potential barriers. This could be circumvented by adding homogeneous laser fields that increase one or both of the inter-potential couplings in a controlled way. This would greatly facilitate fundamental studies of the quantum transport properties in the system. Separating the control of potential depth, dissipation and coupling could then be carried further by detuning so far that the inherent dissipation becomes a small perturbation. In this case, the atomic dynamics cannot be fully described in a classical picture. Even dissipation could be added to the system independently by using standard laser-cooling techniques~\cite{Chu:1998,Cohen-Tannoudji:1998,Phillips:1998}. Therefore, this system may also be relevant in the context of quantum BM's~\cite{Reimann:1997,Hanggi:2005_AnnPh,Hanggi:2005_Chaos}.

%%%%%%%%%%%%%%%%%%%%%%%%%%%%%%%%%%%%%%%%%%%%
%%%%%%%%%%%%%%%%%%%%%%%%%%%%%%%%%%%%%%%%%%%%

In summary, we have demonstrated a 3D BM acting in spatially and temporally symmetric potentials. The need for spatio-temporal asymmetry, which is under-pinned in the Curie principle, results from unequal transition rates between two space-shifted symmetric potentials, thus breaking detailed balance. This has been realized using
ultra-cold atoms in double OL's. A directed drift at constant velocity is demonstrated both vertically and horizontally. By controlling the transition rates and the spatial shift between the potentials, we are able to control the drift velocity in three dimensions. 

The system that we have described in this Letter provides a new platform for general  studies of dynamics in Brownian rectifiers. It is more versatile than other realizations of BM's since it works with symmetric potentials and its parameters can be accurately controlled. Such investigations may have useful applications in general understanding of biological or chemical BM's. Indeed, this mechanism~\cite{Sanchez-Palencia:2004} is not restricted to OL's, but is quite general. The two states could in principle be two different chemical compounds and the transitions could be corresponding chemical reactions~\cite{Astumian:1997}.

\begin{acknowledgments}
This work has been supported by Knut och Alice Wallenbergs stiftelse, Carl Tryggers stiftelse, Kempestiftelserna, Magnus Bergwalls stiftelse and the Swedish Research Council.
\end{acknowledgments}

%\bibliography{../../../../../AK_refs_BM,../../../../../AK_refs_lattice,../../../../../AK_refs_QI,../../../../../AK_refs_general,../../../../../AK_refs_LC,../../../../../AK_refs_own}
%\bibliographystyle{apsrev}

\end{document}